\def\finalpaper{1} % use 1 for final and 0 for anonymous submission

\documentclass[conference]{IEEEtran}
\IEEEoverridecommandlockouts
\usepackage{dblfloatfix}
\usepackage{balance}
\usepackage{tikz}
\usepackage{amsmath}
\usepackage{amsfonts}
\usepackage[ruled,linesnumbered]{algorithm2e}
\usepackage{filecontents}
\usepackage{cuted,adjustbox}
\usepackage{multirow,multicol}
\usepackage{booktabs,colortbl}
\usepackage{pgfplots} 
\usetikzlibrary{patterns}
\usepackage{subcaption}
\usepackage{rotating}
\usepackage{xcolor}
\definecolor{rwth_blue}{RGB}{0,84,159}
\usepackage{circledsteps}
\usepackage{xspace}
\usepackage{tcolorbox}
\tcbuselibrary{listings}
\newcommand\myCircled[2][]{\ifmmode
\Circled[fill color=black,inner color=white,#1]{\mathsf{#2}}
\else
\Circled[fill color=black,inner color=white,#1]{\sffamily#2}
\fi
}
\newtcblisting{foo}{
  listing only,
  nobeforeafter,
  after={\xspace},
  hbox,
  tcbox raise base,
  fontupper=\ttfamily,
  colback=lightgray,
  colframe=lightgray,
  size=fbox
  }
  
\usepackage{listings}
\definecolor{mGreen}{rgb}{0,0.6,0}
\definecolor{mGray}{rgb}{0.5,0.5,0.5}
\definecolor{mPurple}{rgb}{0.58,0,0.82}
\definecolor{backgroundColour}{rgb}{0.95,0.95,0.92}
\lstdefinestyle{CStyle}{
    backgroundcolor=\color{backgroundColour},   
    commentstyle=\color{mGreen},
    keywordstyle=\color{magenta},
    numberstyle=\tiny\color{mGray},
    stringstyle=\color{mPurple},
    basicstyle=\footnotesize,
    breakatwhitespace=false,         
    breaklines=true,                 
    captionpos=b,                    
    keepspaces=true,                 
    showspaces=false,                
    showstringspaces=false,
    showtabs=false,                  
    tabsize=2,
    language=C
}
\usepackage{background}
\usepackage[usestackEOL]{stackengine}
\setstackgap{L}{\normalbaselineskip}
\SetBgContents{\color{blue}{\tiny \Longstack{PREPRINT - accepted at 31st IFIP/IEEE Conference on Very Large Scale Integration (VLSI-SoC 2023).}}}% Set contents
\SetBgPosition{4.5cm,1cm}% Select location
\SetBgOpacity{1.0}% Select opacity
\SetBgAngle{0}% Select rotation of logo
\SetBgScale{1.8}% Select scale factor of logo

\usepackage{array}
\newcolumntype{x}[1]{>{\centering\arraybackslash\hspace{0pt}}p{#1}}

\AtBeginDocument{%
  \providecommand\BibTeX{{%
    \normalfont B\kern-0.5em{\scshape i\kern-0.25em b}\kern-0.8em\TeX}}}

\pgfplotsset{compat=1.18}
\begin{document}
\bstctlcite{IEEEexample:BSTcontrol}

\if\finalpaper1
\title{SoftFlow: Automated HW-SW Confidentiality Verification for Embedded Processors\vspace{-0.25cm}}

\author{Lennart M. Reimann\IEEEauthorrefmark{1}, Jonathan Wiesner\IEEEauthorrefmark{1}, Dominik Sisejkovic\IEEEauthorrefmark{2}, Farhad Merchant\IEEEauthorrefmark{4}, and Rainer Leupers\IEEEauthorrefmark{1}\\
\IEEEauthorrefmark{1}RWTH Aachen University, Germany, 
\{lennart.reimann, wiesner, leupers\}@ice.rwth-aachen.de\\
\IEEEauthorrefmark{2}Corporate Research, Robert Bosch GmbH, Germany, dominik.sisejkovic@de.bosch.com \\
\IEEEauthorrefmark{4}Newcastle University, UK, farhad.merchant@newcastle.ac.uk\\
\vspace{-0.9cm}
}

\else
\title{SoftFlow: Automated HW-SW Confidentiality Verification for Embedded Processors\vspace{-0.25cm}}

\author{Anonymous Authors\\Affiliation\\Affiliation\\Affiliation\\\vspace{-0.9cm}}

\fi

\vspace{-0.6cm}%} % end author
\maketitle

\begin{abstract}
Despite its ever-increasing impact, security is not considered as a design objective in commercial electronic design automation (EDA) tools. This results in vulnerabilities being overlooked during the software-hardware design process.
Specifically, vulnerabilities that allow leakage of sensitive data might stay unnoticed by standard testing, as the leakage itself might not result in evident functional changes. Therefore, EDA tools are needed to elaborate the confidentiality of sensitive data during the design process. However, state-of-the-art implementations either solely consider the hardware or restrict the expressiveness of the security properties that must be proven. Consequently, more proficient tools are required to assist in the software and hardware design. To address this issue, we propose \textit{SoftFlow}, an EDA tool that allows determining whether a given software exploits existing leakage paths in hardware. Based on our analysis, the leakage paths can be retained if proven not to be exploited by software. This is desirable if the removal significantly impacts the design's performance or functionality, or if the path cannot be removed as the chip is already manufactured.
We demonstrate the feasibility of SoftFlow by identifying vulnerabilities in OpenSSL cryptographic C programs, and redesigning them to avoid leakage of cryptographic keys in a RISC-V architecture.
\end{abstract}

\begin{IEEEkeywords}
confidentiality, property checking, information flow analysis, risc-v
\end{IEEEkeywords}

\section{Introduction}

Malicious modifications or unintended insecure software and hardware implementations must be detected at early design stages to avoid expensive post-silicon patches. 
Therefore, Electronic Design Automation (EDA) tools must consider not only performance, power, and area objectives, but also the security implications of hardware and software. Commercial approaches to running secure kernels on secured hardware have already been developed, whereby leakage paths are not considered~\cite{dodo_kernel}. Moreover, the academic community continues to introduce security-aware EDA tools~\cite{cadforassurance}.

\begin{figure}[t!]
    \centering
    \includegraphics[width=\columnwidth]{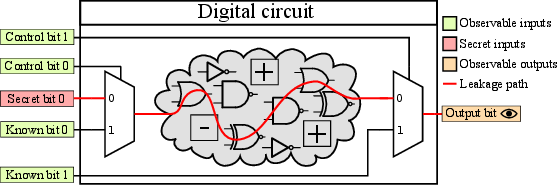}
    \caption{A leakage path in a circuit through a multiplexer.\vspace{-0.45cm} }
    \label{fig:leakage_path}
\end{figure}

Information Flow Analysis (IFA) gains approval when analyzing software or hardware for the leakage of secret data~\cite{rtlift}. However, existing IFA tools do not consider the bare-metal software running on a processor when analyzing the information flow, thus giving \textit{oversensitive} results. Leakage paths that might not be exploited by software are presented as dangerous. Although it is desirable to remove all leakage paths to untrusted components in hardware, it might not always be possible to do so. Sometimes the vulnerabilities are only detected after the manufacturing process, such as Meltdown~\cite{meltdown}. Moreover, removing all leakage paths might result in the loss of functionality or performance. A leakage path is a signal route that carries data from a sensitive source to untrusted components (Fig.~\ref{fig:leakage_path}). 
Fundamentally, there are two distinct flavors of information flow analysis: dynamic and static IFA. A dynamic analysis tracks the information flow within hardware for a given software running a set of test cases~\cite{dift}. As only the program flow for the test cases is considered, vulnerabilities might be overlooked. Thus, static approaches that consider software \textit{independent} of test cases yield more reliable verification results. Most state-of-the-art tools require manual translation of either software or hardware descriptions, or limit the expressiveness of security properties. Thus, a new EDA tool is required to overcome these constraints.
To address this challenge, we present \textit{SoftFlow}, a framework that allows a \textit{hardware-software co-verification} to analyze whether determined leakage paths are activated for a given software and arbitrary input data. 
SoftFlow allows the designer to adapt the program and avoid leakage, or deploy software patches for manufactured chips. 

The contributions of this paper are:
\textbf{(1)} An automated tool to convert leakage paths of a hardware description into provable hardware properties. \textbf{(2)} A complete framework that guarantees that determined leakage paths in hardware are not exploited by a given software. \textbf{(3)} An analysis of the state-of-the-art OpenSSL cryptographic algorithms to demonstrate the usability of the tool for a RISC-V architecture.
%The rest of this paper is organized as follows. Related work is given in Section~\ref{ch:related_work}. Section~\ref{ch:preliminaries} provides background on the threat model, detection of leakage paths in hardware, and the property specification language used in this work. SoftFlow's functionality is elaborated in Section~\ref{ch:softflow} and evaluated in Section~\ref{ch:evaluation}. The results are discussed in Section~\ref{ch:discussion}. Finally, Section~\ref{ch:conclusion} concludes the paper.
\section{Related Work}
\label{ch:related_work}
 \begin{figure*}[!t]
	\centering
	\includegraphics[width=1\textwidth]{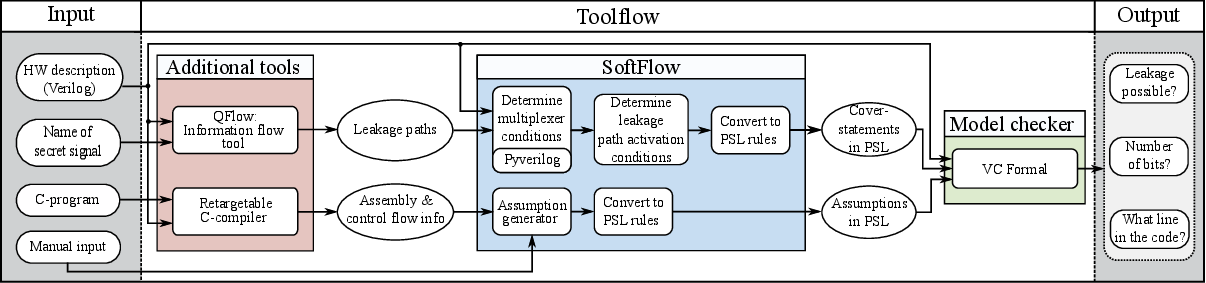}
	 \caption{\label{fig:softflow} Block diagram illustrating \textit{SoftFlow's} tool flow. \vspace{-0.45cm} }
\end{figure*}  
\begin{table*}[hb!]
\vspace{-0.3cm}
    \caption{\label{tab:SERE_operators}Sequential Extended Regular Expression (SERE) operators~\cite{sere_book} used in this work.\vspace{-0.1cm}}
    \centering
    \resizebox{\textwidth}{!}{
    \footnotesize
    \begin{tabular}{c||c|c|l}
         \textbf{Name} &   \textbf{PSL Usage}&  \textbf{Verilog Example} &  \textbf{Interpretation}\\\hline\hline
         \multirow{2}{*}{\textit{CONCAT}} & \multirow{2}{*}{A;B} & \multirow{2}{*}{$(cntrl\_sig1 == 0x20);(cntrl\_sig2 == 0x11)$} & Two cycles exist, so that first condition A is true, followed  \\
         & & & by the cycle in which B is true. \\\hline
         \textit{FUSE} & A:B & $(cntrl\_sig1 == 0x20):(cntrl\_sig2 == 0x11)$ & A cycle exists in that both Boolean expressions A and B are true. \\\hline
         \textit{REPETITION\_INF} & A[*] & $(cntrl\_sig1 == 0x20)[*]$ & The Boolean expression is true in every cycle. \\
    \end{tabular}}
\end{table*}
As the complete system behavior can only be modeled when considering hardware and software together, a co-verification is used frequently to enforce security properties for a given design~\cite{lugou}. 
However, co-verification is a challenging task, as hardware and software are described in different domains.
Typically, one description is either converted into the other domain, or both descriptions are converted to allow the verification of both implementations~\cite{ila_verification}. 
These methods mainly depend on model checkers that formally verify the defined properties. In \cite{soc_trust_jin}, a manual translation of the hardware description, program, and properties is conducted to allow a co-verification. However, a manual translation is error-prone and could introduce vulnerabilities into the description. A model checker, similar to the one used in this work, is used in \cite{interval_property_kunz}\cite{hw_sw_co_drechsler}. However, both proposals limit their use in the variability of the observed software and the expressiveness of the security property. 
With the introduction of SoftFlow, we overcome the mentioned limitations and establish a novel framework that allows an automated generation of security properties. Therefore, we enable a static security analysis of both hardware and software to \textit{facilitate the design of secure software for insecure hardware}.

\section{Preliminaries}
\label{ch:preliminaries}
\subsection{Threat Model} 
The threat model is based on the following assumptions:
\begin{itemize}
    \item The vulnerability that exposes the secret is already present at the RTL-design stage of embedded processors.
    \item The adversary tries to gain information about secret signals, such as cryptographic keys.
   % \item The attack occurs after the processor's manufacturing.
    \item The complete hardware design is known to the attacker.
    \item The adversary can observe primary inputs and outputs of the design.
    \item The software cannot be exchanged by the attacker. 
    \item Arbitrary input data can be used.
\end{itemize}

\subsection{Information Flow Analysis}
IFA allows the static or dynamic analysis of programs and hardware descriptions to elaborate on whether unintended information flow can occur. Security labels are used to flag components to be part of certain \textit{security classes}~\cite{oberg_glift}. Thus, an analysis determines whether sensitive information can be leaked to untrusted areas of the hardware. For the example in Fig.~\ref{fig:leakage_path}, it is analyzed whether the secret bit can be leaked to the observable output for all possible known input bit sequences.

State-of-the-art tools can conduct a static analysis of the Verilog hardware description to determine unwanted data leakages. QFlow~\cite{reimann2021qflow, qflow2} and QIF-Verilog~\cite{qif_verilog} use quantitative information flow to classify data leakages as negligible or a threat. %The amount of leaked information is quantified and compared to a threshold. 
A \textit{scopechain} is provided that allows identifying the path within the hardware description. %, e.g.:
SoftFlow utilizes QFlow to gather the most suspicious leakage path in a system.
\begin{figure*}[!t]
	\centering
	\includegraphics[width=\textwidth]{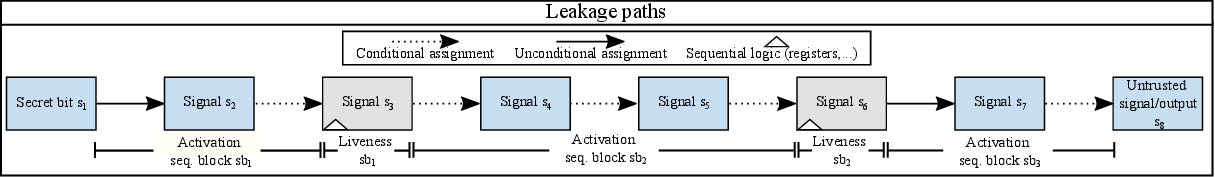}
	\caption{\label{fig:separated_blocks} Separating the leakage paths into blocks to allow a suitable environment for property generation. The arrows illustrate signal assignments in which dotted arrows describe conditional assignments.\vspace{-0.45cm} }
\end{figure*}
\subsection{Property Specification Language (PSL)}
PSL~\cite{practical_psl} is compatible with Verilog hardware descriptions and allows the modeling of the temporal behavior of digital architectures. %~\cite{practical_psl}. 
%Boolean expressions in Verilog can be used to model system behavior. 
A simple example is a condition for a multiplexer:
$(control\_signal == 0x20)$.
The Boolean expression enables the model checker to determine when information is forwarded. 
Moreover, Sequential Extended Regular Expressions (SERE) can model the temporal dependencies. \textit{CONCAT}, \textit{FUSE}, and \textit{REPETITION\_INF} of the SERE repertoire are used in this work, as explained in Table~\ref{tab:SERE_operators}.
The operators can be combined into sequences, allowing the modeling of a leakage path. 
The verification layer uses commands like \texttt{cover} and \texttt{assume}. \texttt{assume} indicates that the included property is expected to hold, thus 
restricting the state-space of the evaluated architecture for the desired property verification. 
However, the user must verify the assumptions to avoid introducing insecurities into the process.
\texttt{cover}-statements can be used to determine whether a property is true for at least one state in the state-space using a model checker. 
If the property is proven to be uncoverable, the leakage cannot occur for the assumptions. SoftFlow automatically generates the assumptions and properties (Fig.~\ref{fig:softflow}).
\begin{figure*}[b]
\centering
\normalsize

\begin{minipage}{\textwidth}
\begin{equation}
\label{eq:path_sequence_example}
\underbrace{\underbrace{\text{\textbf{active(}$s_2$, $s_3$\textbf{)}}}_{AS}\text{ ; }\underbrace{\text{\textbf{alive(}$s_3$\textbf{)}[*]}}_{LS}}_{sb_1}\text{ ; } \underbrace{\underbrace{\text{\textbf{active(}$s_3$, $s_4$\textbf{)} : \textbf{active(}$s_4$, $s_5$\textbf{)} : \textbf{active(}$s_5$, $s_6$\textbf{)}}}_{AS}\text{ ; }\underbrace{\text{\textbf{alive(}$s_6$\textbf{)}[*]}}_{LS}}_{sb_2}\text{ ; }
 \underbrace{\underbrace{\text{\textbf{active(}$s_7$, $s_8$\textbf{)}}}_{AS}}_{sb_3}
\end{equation}
\end{minipage}

\vspace{-0.4cm} 
\end{figure*}
\section{SoftFlow}
\label{ch:softflow}
For the identification of the leakage paths, SoftFlow uses QFlow, due to its capability to analyze the information flow bitwise~\cite{reimann2021qflow}.
QFlow extracts the leakage paths from the sensitive source to the possibly harmful target. 
The security properties are generated by SoftFlow (Fig.~\ref{fig:softflow}), as explained below. 
Additionally, model-checking requires information about the software to constrain the model, reducing the number of false positives. Below, we clarify how the PSL rules are generated, verified and evaluated for multiple programs.
%\subsection{Development Environment}
%The evaluation is conducted with the Synopsys ASIP Designer~\cite{asipdesigner}, which allows a software/hardware co-design of application-specific instruction-set processors. 
\subsection{Property Generation}
This subsection describes a tool that processes the leakage paths, given by a static quantitative IFA, and the hardware description to generate properties using PSL. %Pyverilog~\cite{pyverilog} is used to analyze the leakage path emitted by QFlow. The tool yields information about the structure of this path, which can be used to analyze the condition that needs to be fulfilled to activate the path.

\paragraph{Combinational} First, a combinational circuit is considered. In this case, a leakage path can consist of combinational operations and signal assignments (Fig.~\ref{fig:leakage_path}). These assignment types can be divided into two classes: conditional and unconditional. Unconditional assignments happen independently of any control signals, while conditional assignments depend on a multiplexer's control signal. When checking whether a leak can occur, one needs to analyze whether all conditional assignments are active \textit{at the same time}. 
%For a leakage path with two conditional assignments a cover-statement might look like: \texttt{cover((control\_signal1 == 1) \&\& (control\_signal2 == 2))}. The tools check whether it is possible that \texttt{control\_signal1} and \texttt{control\_signal2} can have the depicted values at the same time. 
Hereafter, \textit{active-function} $A$ describes whether an information flow between two signals is active.
\paragraph{Sequential Blocks} Due to the timing dependency, additional system behavior needs to be considered for sequential circuits. Fig.~\ref{fig:separated_blocks} illustrates a single leakage path. The path leads from a trusted signal, storing a secret bit, to an untrusted signal. %This chain consists of combinational operations, signal/register assignments, and conditional assignments. 
Due to the timing behavior of sequential logic within this path, each leakage path needs to be separated into multiple blocks. 
%This happens at the beginning of the property generation. 
Assignments between non-sequential signals of the same block must occur in a single cycle. But the data can live for several cycles in sequential signals, such as registers, before moving on. Separating the leakage paths into blocks that always end with a sequential signal allows for a differentiation of this behavior.
The separation into the sequential blocks ($sb$) is depicted in Fig.~\ref{fig:separated_blocks}. 
%The leakage path starts at $s_1$ and leaks to the untrusted $s_8$, passing two registers on the way, thus forming three sequential blocks in which the first two end with sequential logic. 
Additionally, one must consider that the secret data is not overwritten within the sequential logic before forwarding it to the next $sb$. 

%For the signal assignments, we can differentiate between conditional and unconditional assignments. 
Only conditional assignments influence the property generation. 
For the first sequential block in Fig. ~\ref{fig:separated_blocks}, only the second assignment holds a condition. Hence, the \textit{block active-function} $BA$ for this sequential block $sb_1$ is true if the conditions for the assignment are true in a single cycle.
\subsubsection{Leakage Path Activation}
First, the sequential logic within the leakage path is identified. This is done to form the sequential blocks. %Consider the three sequential blocks in the given example (Fig.~\ref{fig:separated_blocks}). If the transition from the secret bit $s_1$ up to $s_4$ is active in a single cycle, the data does not reach $s_4$. Due to the delay cycle incurred by a sequential element, the data is only transferred up to $s_3$. Thus, t
The secret datum can only pass a single sequential block in each cycle. For leakage to occur, the second sequential block needs to be activated after the previous one (Fig.~\ref{fig:separated_blocks}). 
Additionally, the two blocks do not have to be activated in consecutive cycles
as long as the information in the previous sequential logic stays alive and is not overwritten until it passes the next sequential block, which is described with the \textit{alive-function} $L$.
%This liveness needs to be covered in the properties as well. 
%No secret may be overwritten until it has been passed to the next sequential block. %which is described by the \textit{alive-function} $L$.
%For a leakage path to leak data, the different conditional assignments need to be true.
%Additionally, the sequential logic at the end of a sequential block needs to hold the data until it is forwarded to the next register, which is described by the \textit{alive-function} $L$.
Each sequential block results in a $BA$ and $L$ function, except for the last one. 
%These functions need to be true in subsequent cycles for the consecutive sequential blocks. 
In the example (Fig.~\ref{fig:separated_blocks}), two alive-functions and three block active-functions are required to describe the activation of the path.
%As only five signal assignments are conditional, five active-functions are required. 
\subsubsection{Active \& Alive-Function Collection}
%Algorithm~\ref{alg:activation_condition} generates the PSL expressions for the transition activations. The input consists of the set of binds that contains the two signals and the name of the two signals. The algorithm yields the activation condition for its transition. 
% \setlength{\textfloatsep}{10pt}
% \begin{algorithm}[!t]
% 	\SetAlgoLined
% 	\footnotesize
% 	\KwIn{T($s_i$, $s_{i+1}$), set of binds B}
% 	\KwOut{$activeCon$ for the given T}
% 	$activeCon$ $\leftarrow$ FALSE\\
% \Begin{
% 	$assignmentPathsList$ $\leftarrow$ \textbf{getPathsOfBind(}B($s_{i+1}$), $s_i$\textbf{)}\\
% 	\For{assignmentPath in assignmentPathsList}{
% 		$pathCon$ $\leftarrow$ TRUE\\
% 		$tree$ $\leftarrow$ \textbf{getRoot(}B($s_{i+1}$)\textbf{)}\\
% 		\For{edge E in assignmentPath}{

% 			\If{$\tau_E$(E) in \{TRUE, FALSE\}}{
% 			    connect $\leftarrow$ \textbf{getConnect(}$tree$, COND\textbf{)}\\
% 				$branchCon$ $\leftarrow$ \textbf{getCondExpress(}connect\textbf{)}\\
% 				$pathCon$ $\leftarrow$ $pathCon$ $\land$ ($branchCon$ == $\tau_E$($E$))
% 				}
% 			$tree$ $\leftarrow$ \textbf{getConnect(}$tree$, $E$\textbf{)}
% 		}
% 		$activeCon$ $\leftarrow$ $activeCon$ $\lor$ $pathCon$
% 	}
% }
% 	\caption{\label{alg:activation_condition}Extracting the activation condition}
% \end{algorithm}
First of all, the Verilog description is parsed to form a tree structure that can be iterated over.
%Pyverilog~\cite{pyverilog} is used to parse the hardware description in Verilog. 
%The parser produces a tree of binds that is processed into the individual active-functions $A$ for each conditional signal assignment, which is stored in $activeCon$ (line 15). The function \textbf{getPathsOfBind} yields all possible assignment paths through the binds $B(s_{i+1})$ of the destination signal $s_{i+1}$ from the source $s_i$. 
Afterward, the algorithm iterates over the leakage paths to determine the individual activation conditions. The conditions for a single signal assignment are connected via an OR-operator as only a single true activation condition can lead to leakage. The resulting set of active-functions are temporally stored in a file to be processed to PSL.
%For each \textit{assignmentPath} in \textit{assignmentPathsList} (line 4) the root of the tree is determined, which represents the node for the destination signal. Afterward, the algorithm iterates over the different edges of the current assignment path. For conditional assignments (indicated by \textit{TRUE} or \textit{FALSE}) the node is returned by \textbf{getConnect}. The conditional expression is stored in \textit{branchCon} using \textbf{getCondExpress}. Combining it with the branch condition $\tau_E$(E) results in the final \textit{pathCon} of this edge. All \textit{pathCon} are connected to the complete condition for that assignment path. The activation conditions are determined for every sequential block consisting of all possible transfers of information from one signal to the next. 
%\textcolor{red}{REWRITE}
%\subsubsection{Implementing the Alive-Function}
%\label{sec:impl_liveness}
The alive-function $L$ is determined similarly to the active-function. 
%For this purpose, Algorithm~\ref{alg:activation_condition} can be reused. 
The alive-function returns 'true' if the current value in a sequential logic is not overwritten for a fixed amount of cycles. An empty signal assignment describes a value that is not overwritten. 
%In the bind tree it is represented by an assignment from an empty node to the destination signal, the sequential logic. 
The activation condition from the empty node to the destination node is calculated and stored as the alive function. 
%As long as these conditions are true, the value within the register or latch stays valid.
%This algorithmic function for a signal S can then be denoted as $valid(S)$, representing the alive-function. 
%If the function needs to be evaluated at a special cycle c during the runtime, the function $valid_c(S)$ is used. 
\subsubsection{Specifying SoftFlow's PSL Sequences}
%The discussed algorithms outline all required functionalities to describe the activation of a leakage path in a sequential circuit. 
The collected active-functions and alive functions need to be translated to PSL to enable model-checking. For the example in Fig.~\ref{fig:separated_blocks}, the PSL sequence is shown in Equation~\ref{eq:path_sequence_example}, with $AS$ and $LS$ describing the activation and liveness sequences.
%The \textbf{active} and \textbf{valid} commands are replaced with their Boolean equivalent 
%that allow the leakage of information.
The sequence is extended with the \texttt{cover} command, forming the \textit{cover rules}. If the model checker can cover this path, the leakage can occur for the given assumptions. 
%The translation is depicted in Algorithm~\ref{alg:psl_sequence}.
%\setlength{\textfloatsep}{0cm}% Remove \textfloatsep
%  \begin{algorithm}[t]
%  	\SetAlgoLined
%  	\footnotesize
%  	\KwIn{a leakage path description $lPath$}
%  	\KwOut{a PSL sequence describing the path activation $pathSeq$}
%  	\Begin{
%  		$SB$ $\leftarrow$ \textbf{genSBs(}$lPath$\textbf{)}\\
%  		$sbSeqList$ $\leftarrow$ empty list\\
%  		\For{$sb$ in SB}{
%  			$transitConList$ $\leftarrow$ empty list\\
%  			\For{t($s_i$, $s_{i+1}$) in $sb$}{
% 				$transitCon$ $\leftarrow$ \textbf{active(}t($s_i$, $s_{i+1}$)\textbf{)}\\
%  				\textbf{append} $transitiCon$ to $transitConList$\\
%  			}
%  			$transitSeq$ $\leftarrow$ \textbf{transitConToPSL(}$transitConList$\textbf{)}\\
%  			%\If {sb \text{is not last in} SB}{
%  			$livenessCon$ $\leftarrow$ \textbf{valid(}$sb_{SL}$\textbf{)}\\
%  			$livenessSeq$ $\leftarrow$ \textbf{livenessConToPSL(}$livenessCon$\textbf{)}\\%}
%  			$sbSeq$ $\leftarrow$ \textbf{combine(}$transitSeq$, $livenessSeq$\textbf{)}\\
%  			\textbf{append} $sbSeq$ to $sbSeqeList$\\
%  		}
%  		$pathSeq$ $\leftarrow$ \textbf{combine(}$sbSeqList$\textbf{)}\\
%  	}
%  	\caption{\label{alg:psl_sequence}Generating a PSL sequence}
%  \end{algorithm}
% First of all, the leakage path is separated into sequential blocks using \textbf{genSBs}. 
 %Afterward, the PSL conditions for each sequential block are determined iteratively using \textbf{active}, which represents Algorithm~\ref{alg:activation_condition}.  
The individual Boolean activation conditions are connected using the \textit{FUSE} operator (Table~\ref{tab:SERE_operators}), as the conditions must be true in the same cycle. 
%For $n$ conditional assignments, this yields:
%\begin{equation*}
%\text{\textbf{transitionConToPSL(}$b_1$, $b_2$, ..., $b_n$\textbf{)}} = b_1 : b_2 : ... : b_n
%\label{eq:transition_fusion}
%\end{equation*}
%No fusion-operator is required for the first SB in Fig.~\ref{fig:separated_blocks}, as only a single conditional assignment is present. 
After the activation conditions are converted, the liveness condition for the sequential logic is required. The returned liveness condition is converted to a PSL expression using the \textit{REPETITION\_INF} operator (Table~\ref{tab:SERE_operators}).
%, as illustrated in the equation below:
%\begin{equation*}
%\text{\textbf{livenessConToPSL(}$b$\textbf{)}} = b[*]
%\label{eq:liveness_sequence}
%\end{equation*}
The activation conditions and the alive-functions are combined into the full sequence for a single $sb$ using \textit{CONCAT} (Table~\ref{tab:SERE_operators}).
%function for $n$ sequential blocks, given the individual activation sequences $sq$: 
%\begin{equation*}
%\text{\textbf{combine(}$sq_1$, $sq_2$, ... ,$sq_n$\textbf{)}} = sq_1 ; sq_2 ; ...; sq_n
%\label{eq:combine}
%\end{equation*}
%The \textbf{combine} function uses the concatenation operator, which allows combining properties into sequences. 
%In this case, the operator states that $sq_2$ follows $sq_1$ in the consecutive cycle. 
\textit{Uncoverable properties state that the leakage path cannot be taken}. 
\subsection{Property Verification}

By employing model-checking, one can establish a level of \textit{certainty} regarding the reliability of the verification results. %Thus, we focus on the formal methodology instead of simulation. 
%The inputs should be constrained as much as possible to avoid false positives. 
The individual cover rules allow an independent evaluation for every leakage path, enabling the model checker to work in parallel for these properties.
%VC Formal ~\cite{vcformal} is used to apply the model-checking using the generated PSL rules. 
%The model checker outputs the state of the architecture for which the properties are proven to be covered. 
The program memory address is used to identify the instruction and the related C code line. % using ASIP Designer.

SoftFlow's generated properties do not consider the compiler program for this processor. As we would like to elaborate on the security of a given C program and an insecure hardware implementation, the hardware model for the model-checking needs to be constrained before running the verification. This is done to reduce the false positives, which describe a leakage of secret data for programs that would not be implemented. Those constraints are implemented using \texttt{assume}. 
Since the model permits the data memory to contain any information, it is necessary to contemplate all conceivable routes within a program. Conditional branches depend mostly on data in the data memory. Thus, all possible conditional jumps need to be considered. The user can choose from the following derived assumptions (\myCircled{A} to \myCircled{F}) to restrict the program flow.
%As a reminder, if any of the used assumptions are false for the given software and hardware, \textit{insecure hardware is labeled as secure.}

\emph{\myCircled{A} No illegal instructions:}
% compiler yields a file that describes the unused instruction encodings for the processor model. 
A assumption is made which claims that the read port of the program memory can never hold the value of prohibited instructions.

\emph{\myCircled{B} Only used instructions:}
%At this point, the actual compiled program is considered. 
The formal model can be further constrained by analyzing the compiled machine-code statically. An assumption can be generated that further constrains the model by stating that only instructions present in this compiled program can be read from the program memory. The order is not yet considered.

\emph{\myCircled{C} Replacing the program memory:}
%Now, the order of the program flow is considered. 
To avoid modeling an entire program memory, a lookup table is used instead, which also enables consideration of the order of instructions.
Furthermore, since return addresses of function calls are stored on the stack, it is possible for a return to transpire to any point within the program.

\emph{\myCircled{D} Only legal return addresses:}
The remaining three assumptions are generated using compiler information.
The assumption is used to remove undefined return addresses, which results in arbitrary program flows for the verification. 

\emph{\myCircled{E} Only correct hardware jumps:}
Compiler information is processed to allow only valid start and return addresses for hardware loops during the evaluation.
%For hardware loops internal control registers are used to store, e.g., the start and end addresses of a loop. The control registers are filled at runtime with data. 
%This results in a similar issue as with the return addresses. 

\emph{\myCircled{F} Call-return matching:}
When a function is invoked from several locations, it is essential to account for both return addresses. Confirming a legitimate correlation between a call and its corresponding return is only achievable during runtime. Nonetheless, if an authentic hardware call-stack is utilized to store the return addresses, it can already be taken into consideration during the verification process.
\section{Evaluation}
\label{ch:evaluation}
\begin{figure}[!t]
    \centering
    \includegraphics[width=0.99\columnwidth]{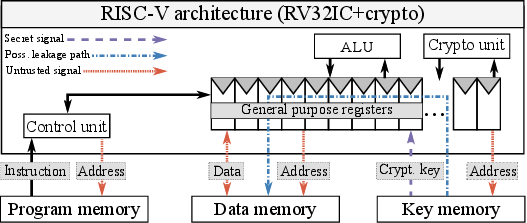}
    \caption{Abstract block diagram of the RISC-V architecture.\vspace{-0.4cm} }
    \label{fig:riscv}
\end{figure}

The evaluation is conducted for a RISC-V Verilog description and compiler.
%model is used to generate a compiler and a Verilog description. 
The processor uses the RV32IC instruction-set, in which an instruction for a single round of AES, and an additional memory have been added that holds the cryptographic key. 
A block diagram illustrating the architecture is shown in Fig.~\ref{fig:riscv}. The data input from the key memory is labeled \textit{sensitive} in order for QFlow to yield the most suspicious leakage paths from the signal to all output ports of the design, including the memories' data and address ports. 
%The leakage paths are delivered by a static IFA tool. 
The analysis is performed for several cryptographic algorithms from OpenSSL\cite{openssl}: ChaCha20, AES-256, Camelia, Aria, and SHA-256. 
SoftFlow is evaluated for different assumptions to elaborate on their efficiency and security.
\subsection{Verification Cases}
The used assumptions are listed behind each mode.

\underline{\textsc{None}:}
No assumptions are added to the model checker. The verification tool works solely with the properties generated from our property generator for the leakage paths. 

\underline{\textsc{Legal} \myCircled{A}:}
The loading port of the program memory is restricted, disallowing all illegal instructions to be read.

\underline{\textsc{Used} \myCircled{A}\& \myCircled{B}:}
The data port of the program memory is restricted using an \texttt{assume} statement that allows only instructions present in a compiled machine-code to be read. %via the port.

\underline{\textsc{Jumps} \myCircled{A} to \myCircled{E}:}
In this mode, only valid return addresses can be used for a return command and hardware loops, allowing a more realistic program flow.

\underline{\textsc{Stack} \myCircled{A} to \myCircled{F}:}
A hardware stack allowing the return-call matching is added as supplemental Verilog code.

\underline{\textsc{Full}:}
First, the assumptions for \textsc{Used} are utilized. If the property for this path cannot be proven to be uncoverable, all verification cases are tried until the model checker is successful. 
Considering the valid states S of a processor model, the valid states for \textsc{Stack} are present in \textsc{Used}, resulting in:
\begin{equation*}
S_{\textsc{Stack}}\subseteq S_{\textsc{Jumps}}\subseteq S_{\textsc{Used}}\subseteq S_{\textsc{Legal}}\subseteq S_{\textsc{None}}
\end{equation*}
The inclusiveness of the more restrictive verification cases further reduces the state space that has to be analyzed.
%The non-activation of leakage paths can already be proven for some of the given paths with an extensive set of valid processor states. However, more restrictions commonly lead to higher runtimes.
\subsection{Results}
\subsubsection{Coverages}
The evaluation results of SoftFlow's functionality are presented below. The verification modes are applied to the five cryptographic algorithms for the RISC-V.
\paragraph{Without Software:}
\begin{figure}[!t]
\centering
\begin{tikzpicture}[overlay]
		\draw[thick, -stealth, red] (2.9,2.4) -- (3.5,2.5);
		\draw[dashed, line width=0.8mm, red] (0, 2.15) -- (0, 2.75);
	\end{tikzpicture}
\begin{subfigure}{0.37\columnwidth}
 \begin{lstlisting}[basicstyle=\footnotesize]
  .
ldk  x7,  1(x6!) 
addi x10, x13, 40 
sw   x13, 0(x2) 
li   x6,  1 
sw   x7,  4(x13!) 
mv   x14, x7 
ldk  x7,  1(x6!) 
sw   x7,  4(x13!) 
mv   x15, x7 
ldk  x7,  1(x6!) 
sw   x7,  4(x13!) 
sw   x7,  4(x2) 
\end{lstlisting}
\caption{\label{fig:naive_app_assembly}Assembly program}
\end{subfigure}
\begin{subfigure}{0.6\columnwidth}
%\lstset{numbers=left}
	\begin{lstlisting}[style=CStyle, basicstyle=\footnotesize]
int chess_storage(KM:0) km[4];
int Camellia_Ekeygen(int keyBitLength,
		     KEY_TABLE_TYPE k)
{
	register u32 s0, s1, s2, s3;
	
	k[0] = s0 = km[0];
	k[1] = s1 = km[1];
	k[2] = s2 = km[2];
	k[3] = s3 = km[3];
	.
	s0 ^= k[0], s1 ^= k[1],\
\end{lstlisting}
\caption{\label{fig:naive_app_c}C program}
\end{subfigure}
\caption{\label{fig:naive_app} A naive implementation of Camellia: The dashed line symbolizes the reported instructions activating the leakage path. The corresponding C-code is marked with an arrow.\vspace{-0.4cm} }
\end{figure}
The assumptions of the modes \textsc{None} and \textsc{Legal} are applied to the verification model. These two modes are independent of the actual software running.
%Thus, no individual elaborations for the algorithms are required. 
QFlow yields \textbf{3776} leakage paths that need to be elaborated for the given architecture.
% \begin{table}[!b]
% \vspace{-0.1cm}
% \caption{\label{tab:stats_none}Statistics of the leakage paths grouped by status covered and uncoverable for the case \textsc{None}.}
% \footnotesize
% 	\centering
% 	\vspace{-0.1cm}
% \begin{tabular}{c||c|c}
% 	 & \multicolumn{2}{c}{\textbf{Leakage}}\\
% 	\textbf{Status}&\textbf{Sum (bit)} &\textbf{Count}\\
% 	\hline \hline
% 	covered&	4.30&		2919\\
% 	uncoverable&	0.27&		857\\
% 	\hline
% 	combined & 	4.57 &  3776\\
% \end{tabular}
% \end{table}
Leakage paths are \textbf{uncoverable} if the leakage cannot occur for the given assumptions. The paths that can be activated are labeled \textbf{covered}. Both, the \textsc{Legal} and \textsc{None} modes yield \textbf{857} uncoverable paths, which indicates that the processor only accepts legal instructions and that the 857 paths are false positives from QFlow. \textit{With SoftFlow, the false positives are removed entirely!}
\paragraph{With OpenSSL Cryptographic Algorithms:}
Next, the capabilities of SoftFlow are elaborated for the different OpenSSL algorithms for the remaining verification modes. 
An example of how the applications are modified for the architecture is shown in Fig.~\ref{fig:naive_app_c}. The key can be accessed with a pointer to the key memory KM. Table~\ref{tab:eval_applications} illustrates the results of the elaboration for the modes \textsc{Stack}, \textsc{Jumps}, and \textsc{Used}. As one can see, the number of uncoverable paths increases for all applications when more restrictive assumptions are used. 
\begin{table}[!b]
\setlength\extrarowheight{3pt}
\vspace{-0.1cm}
\caption{\label{tab:eval_applications} Summarized metrics for the five applications grouped by verification case.\vspace{-0.1cm}}
    \footnotesize
	\centering
	\resizebox{\columnwidth}{!}{
	\vspace{-0.2cm}
	\begin{tabular}{p{0.02cm}c||c|c|c|c|c}
		&\textbf{Application}& \textbf{Aria} & \textbf{ChaCha20} & \textbf{Camellia} & \textbf{AES} & \textbf{SHA}\\
		\hline \hline
		&Instr. count&2112 & 1347 & 1973 & 1409 & 649\\
		\hline
		&Call depth & 4 & 4 & 4 & 6 & 5\\
		\hline
		\multirow{2}{*}{\begin{turn}{90}\textsc{Used}\end{turn}} 
		& Covered & 65 & 74 & 58 & 58 & 48\\
		&Uncoverable & 3711 & 3702 & 3718 & 3718 & 3728\\
	%	&$L_{\Sigma}$ (bit) & 2.47 & 2.74& 2.46 &2.46 & 1.36 \\
		\hline
		\multirow{2}{*}{\begin{turn}{90}\textsc{Jumps}\end{turn}} 
		& Covered & 32 & 32 & 32 & 32 & 32\\
		&Uncoverable & 3744 & 3744 & 3744 & 3744 & 3744\\
	%	&$L_{\Sigma}$ (bit) & 1.07 & 1.07 & 1.07 & 1.07 & 1.07\\
		\hline
		\multirow{2}{*}{\begin{turn}{90}\textsc{Stack}\end{turn}} 
		& Covered & 32 & 32 & 32 & 32 & 32\\
		&Uncoverable& 3744 & 3744 & 3744 & 3744 & 3744\\
	%	&$L_{\Sigma}$ (bit)& 1.07 & 1.07 & 1.07 & 1.07 & 1.07\\
	\end{tabular}}
\end{table}
No difference in the results between the modes \textsc{Stack} and \textsc{Jumps} can be observed.

The call tree depths are limited for all five applications, which is required if the \textsc{Stack} mode is used. Otherwise, the hardware stack would have to be of unlimited size. Thirty-two leakage paths are still covered for the design and applications. The cause for this leakage detection can be seen in Fig.~\ref{fig:naive_app} for the application Camellia. 
The remaining exemplary elaborations are only presented for Camellia. Final results are presented for all applications. The assembly code in Fig.~\ref{fig:naive_app_assembly} shows that the compiler loads the key values from the dedicated key memory (\textit{ldk}, load key) and stores them in the untrusted data memory (\textit{sw}, store word) for the array \textit{k}. The programmer can avoid the usage of the leakage path by modifying the C program. Listing~\ref{lst:ktokm} shows the replacement of the variable \textit{k} with the pointer \textit{km}, so that the values are not automatically stored in the untrusted data memory. Additionally, the pointer is marked as \textit{volatile} (Listing~\ref{lst:volatile}), instructing the compiler not to optimize any data movements for KM's data. Table~\ref{tab:modified_applications} depicts the outcomes of SoftFlow subsequent to this modification. The leakage paths are not activated anymore except for the Aria application. 
%For all other cryptographic applications, it is proven that none of the \textbf{3776} leakage paths are activated for the applications.
For Aria, the compiler places the key in a register that is directly connected to the address port of the key memory, which is marked as untrusted. Forcing the compiler to pick a different register can also avoid this leakage. However, the limited intrusion into the compiler does not allow this change. The final evaluation shows an example of finding software Trojans inside the program. In the four applications that achieved 100\% avoidance of leakage, a software Trojan is implemented. An example Trojan for the Camellia application can be seen in Listing~\ref{fig:trojan_camellia}. For a particular input value stored in the data memory, the keys are written to the untrusted memory. 
%The assembly program on the left shows how the leakage paths from the key memory to the data memory are triggered, loading the key (\textit{ldk}, load key) and storing it to untrusted components (\textit{sw}, store word). 
\lstset{numbers=left,xleftmargin=2em,framexleftmargin=1.5em}
\begin{figure}[!t]
\begin{lstlisting}[style=CStyle, firstnumber=14, caption=Replacing k with km in the application camellia, label=lst:ktokm]
	s0 ^= km[0], s1 ^= km[1],\
	s2 ^= km[2], s3 ^= km[3];
\end{lstlisting}
\vspace{-0.5cm} 
\end{figure}
\begin{figure}[!t]
\begin{lstlisting}[style=CStyle, caption=Declaring km as \textbf{volatile}, label=lst:volatile]
	volatile int chess_storage(KM:0) km[4];
\end{lstlisting} 
\vspace{-0.8cm} 
\end{figure}\lstset{numbers=left,xleftmargin=2em,framexleftmargin=1.5em}
\begin{figure}[!t]
	\begin{lstlisting}[caption= The malicious lines implementing a simple Trojan in the application Camellia., label=fig:trojan_camellia, style=CStyle]
if (input[0] == 0x4d2)
{
	u32* leakage = (u32*) malloc(sizeof(u32)*8);
	for (int j=0; j<8; j++) leakage[j] = km[j];
}
\end{lstlisting}
\vspace{-0.5cm} 
\end{figure}
\begin{table}[b]
	\vspace{-0.3cm}
	\caption{\label{tab:modified_applications} Metrics of the modified applications.\vspace{-0.1cm}}
\footnotesize
	\centering
\resizebox{\columnwidth}{!}{
	\begin{tabular}{p{0.02cm}x{1.7cm}||x{0.75cm}|x{1.4cm}|x{1.28cm}|x{0.65cm}|x{0.65cm}}
		&\textbf{Application}& \textbf{Aria} & \textbf{ChaCha20} & \textbf{Camellia} & \textbf{AES} & \textbf{SHA}\\
		\hline \hline
		\multirow{3}{*}{\begin{turn}{90}\textsc{Full}\end{turn}} 
		& covered & 473 & 0 & 0 & 0 & 0\\
		&uncoverable & 3303 & 3776 & 3776 & 3776 & 3776\\
	%	&$L_{\Sigma}$ (bit)& 0.22 & 0& 0 & 0 & 0 \\
		&time & 16016 & 16780 & 3213 & 1337 & 4432\\
		
	\end{tabular}}
\end{table}
\begin{table}[b]
\vspace{-0.1cm}
	\caption{\label{tab:modified_applications_trojan} Verification of the modified applications with embedded Trojans.}
	\vspace{-0.1cm}
\footnotesize
	\centering
		\begin{tabular}{p{0.02cm}c||c|c|c|c}
			&\textbf{Application}& \textbf{ChaCha20} & \textbf{Camellia} & \textbf{AES} & \textbf{SHA}\\
		\hline \hline
		\multirow{5}{*}{\begin{turn}{90}\textsc{Full}\end{turn}} 
		& covered & 32 & 32 & 32 & 32\\
		& uncoverable & 3744 & 3744 & 3744 & 3744\\
		%&depth & 2.7521 & 2.7727 & 2.8389 & 2.8050 \\
		&time & 26045 & 2127 & 1386 & 1321\\
		&trigger (hex)& 12345678 & 4d2 & fedcba & 12ef34dc\\
		
	\end{tabular}
\end{table}

The results for the evaluations of the \textsc{Full} mode illustrate \textit{that all Trojans that use the leakage paths are detected despite their difference in the trigger}, as shown in Table~\ref{tab:modified_applications_trojan}. The value of the trigger does not play a role in the detection, as the data coming from the untrusted data memory is assumed to be of any possible value. 
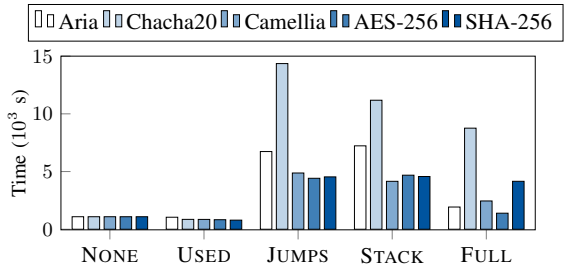
\begin{figure}[t]
    \centering
\begin{tikzpicture}[scale=0.85]
    \begin{axis}[
        ybar,
        enlargelimits=false, 
        scaled y ticks = false,
        %legend style={at={(0.02,0.74)},
        %anchor=west},
        legend style={at={(0.495,1.3)}, anchor=north,legend columns=-1},
        bar width=0.18cm,
        width=9cm,
        height=4.3cm,
        ylabel={Time ($10^3$ s)},
        ymax=15.000,
        ymin=-0.050,
        enlarge x limits=0.13,
        symbolic x coords={\textsc{None},\textsc{Used},\textsc{Jumps},\textsc{Stack},\textsc{Full}},
        xtick=data,
        ylabel near ticks,
        ylabel shift={-0.55em},
        xtick pos=left,
        ytick pos=left,
        legend image post style={scale=0.9},
        y tick label style={font=\small},
        y label style={font=\small}
    ]
    \addplot[fill=white] coordinates {(\textsc{None},1.085) (\textsc{Used},1.041) (\textsc{Jumps},6.733) (\textsc{Stack},7.228)(\textsc{Full},1.925)};
    \addplot[fill=rwth_blue!25] coordinates {(\textsc{None},1.085) (\textsc{Used},0.856) (\textsc{Jumps},14.356) (\textsc{Stack},11.180)(\textsc{Full},8.760)};
    \addplot[fill=rwth_blue!50] coordinates {(\textsc{None},1.085) (\textsc{Used},0.852) (\textsc{Jumps},4.874) (\textsc{Stack},4.153)(\textsc{Full},2.451)};
    \addplot[fill=rwth_blue!75] coordinates {(\textsc{None},1.085) (\textsc{Used},0.826) (\textsc{Jumps},4.416) (\textsc{Stack},4.684)(\textsc{Full},1.391)};
    \addplot[fill=rwth_blue] coordinates {(\textsc{None},1.085) (\textsc{Used},0.789) (\textsc{Jumps},4.537) (\textsc{Stack},4.570)(\textsc{Full},4.155)};

    \legend{Aria, Chacha20, Camellia, AES-256, SHA-256}
    \end{axis}
    \end{tikzpicture}  
    
    \caption{\label{fig:runtime}The summarized runtime of the verification procedure for the different verification modes and applications.\vspace{-0.45cm}}
\end{figure}
\subsubsection{Runtime}
%This section evaluates the runtime cost of SoftFlow. 
For the unmodified applications (e.g., Fig.~\ref{fig:naive_app_c}) that still carry 32 covered paths, the verification is conducted by using the available verification modes. \textsc{Legal} is not presented here, as it yielded the same verification results as \textsc{None}. As shown in Fig.~\ref{fig:runtime}, the lowest runtimes are given by the verification modes that restrict the hardware model the least. However, some leakage paths could not be marked as \textbf{uncoverable} for the less restrictive assumptions (Table~\ref{tab:eval_applications}). 

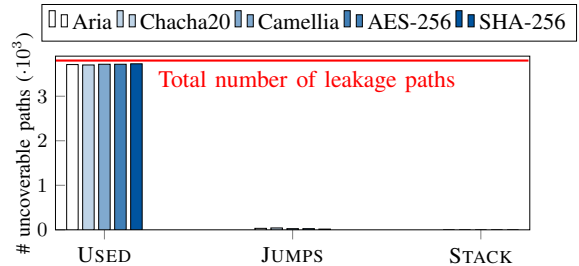
\begin{figure}[t]
    \centering
\begin{tikzpicture}[scale=0.85]
		\draw[thick, red] (0, 2.65) -- node[below, xshift=0.2cm]{\small Total number of leakage paths}(7.4, 2.65) ;%;

    \begin{axis}[
        ybar,
        enlargelimits=false,
        %legend style={at={(0.02,0.74)},
        %anchor=west},
        legend style={at={(0.535,1.3)}, anchor=north,legend columns=-1},
        bar width=0.18cm,
        width=9cm,
        height=4.3cm,
        ylabel={\# uncoverable paths ($\cdot 10^3$)},
        ymax=3.900,
        enlarge x limits=0.13,
        symbolic x coords={\textsc{Used},\textsc{Jumps},\textsc{Stack}},
        xtick=data,
        ylabel near ticks,
        ylabel shift={-0.55em},
        xtick pos=left,
        ytick pos=left,
        legend image post style={scale=0.9},
        y tick label style={font=\small},
        y label style={font=\small}
    ]
    \addplot[fill=white] coordinates {(\textsc{Used},3.711) (\textsc{Jumps},0.033) (\textsc{Stack},0)};
    \addplot[fill=rwth_blue!25] coordinates {(\textsc{Used},3.702) (\textsc{Jumps},0.042) (\textsc{Stack},0)};
    \addplot[fill=rwth_blue!50] coordinates {(\textsc{Used},3.718) (\textsc{Jumps},0.026) (\textsc{Stack},0)};
    \addplot[fill=rwth_blue!75] coordinates {(\textsc{Used},3.718) (\textsc{Jumps},0.026) (\textsc{Stack},0)};
    \addplot[fill=rwth_blue] coordinates {(\textsc{Used},3.728) (\textsc{Jumps},0.016) (\textsc{Stack},0)};
    %\addplot[red,sharp plot]
coordinates {(\textsc{None},3776) (\textsc{Stack},3776)}
;
    \legend{Aria, Chacha20, Camellia, AES-256, SHA-256}
    \end{axis}
    \end{tikzpicture}  
    
    \caption{\label{fig:second_bar} A histogram showing what modes are mostly successful when verifying the model for different applications. \vspace{-0.45cm}}
\end{figure}

The runtime increases drastically for all applications at \textsc{Jumps}. Moreover, as expected, the lowest runtime can be achieved using the verification mode \textsc{Full}. The overall runtime is optimized, while yielding a precise result for all applications.
As the verification of the properties for every leakage path can be parallelized, depending on the available resources of the designer, the runtimes can be drastically reduced.
%If all paths are elaborated in parallel, the highest runtime of ChaCha20 (14356 seconds) would be reduced to about three seconds on average, whereby the runtime of each verification depends on the complexity of the path. 
Fig.~\ref{fig:second_bar} illustrates how successful the different verification modes are in their task. It can be observed that most leakage paths can be flagged as uncoverable by assuming that only the instructions given in a program will be used. 
%The implementation of the hardware stack is not required for any of the given applications but might be useful for more complex applications. 
The higher number of cases verified using \textsc{Used} explains the reduced runtime of the \textsc{Full}-mode in Fig.~\ref{fig:runtime}.
\section{Limitations}
\label{ch:discussion}
Overall, the tool can identify vulnerabilities and assist the designer in writing software that allows safe use of insecure hardware. However, some limitations must be considered.

Leakage paths can be triggered with instructions like load and store operations, as shown in Fig.~\ref{fig:naive_app}. The data is leaked without any change. This results in a varied attack scenario from the one used by QFlow. 
Additionally, the evaluation is only conducted on 3776 existing leakage paths, as QFlow only outputs the most suspicious ones. Although the elaboration of the paths was successful, all leakage paths must be considered. 
It might not be possible for some applications and hardware combinations to avoid leakage. This could be due to unconditional leakages or leakages caused by instruction patterns required for the application.
%The \textsc{Stack} verification mode did not bring any advantage in the evaluation. It might be the case that some applications are easier to verify using this mode. 
 %Furthermore, SoftFlow's verification is only reliable as long as the assumptions hold. 

\section{Conclusion}
\label{ch:conclusion}
We presented SoftFlow, an EDA tool that utilizes formal verification to conclusively indicate whether leakage paths in hardware are activated for a given software. Therefore, SoftFlow facilitates both a security-driven system design from the ground up and post-fabrication security patches. The usability of the proposed framework was demonstrated by analyzing OpenSSL cryptographic algorithms for a RISC-V architecture. The evaluation proves that secure software disabled all 3776 leakage paths. In addition, the consideration of both software and hardware mitigated all false positives of QFlow.
With SoftFlow, we enable a security-aware software-hardware co-verification process that takes into account the intricate interplay of dedicated hardware and software. 
In the future, we plan to investigate security-aware compilers that facilitate the complete removal of vulnerabilities if allowed by both hardware and software.
%Additionally, the leakage paths that are used in this work describe a direct leakage via signals. However, if valid information about the activation of power, EM, and timing side-channels are known via any other tool, a similar property generator can be written to support these channels by SoftFlow as well. Furthermore, leakage paths to peripherals and third-party intellectual property can be handled similarly to the paths in this work. 

%-------------------------------------------------------------------------------
\bibliographystyle{IEEEtran}
\bibliography{bibtexentry}

%%%%%%%%%%%%%%%%%%%%%%%%%%%%%%%%%%%%%%%%%%%%%%%%%%%%%%%%%%%%%%%%%%%%%%%%%%%%%%%%
\end{document}